# Demonstration of non-Abelian frame charge flow in photonic crystals

Dongyang Wang[1], Z. Q. Zhang[1], C. T. Chan[1]

1. Department of Physics and Center for Metamaterials Research, Hong Kong University of Science and Technology; Hong Kong, China.

*Correspondence to: phchan@ust.hk

**Abstract**

In *PT* symmetric systems, the notion of non-Abelian frame charges enables multiband topological characterization of the degeneracy nodes through examining the eigenvector frame rotations. Interestingly, some features of these frame charges can be viewed as an analogue of electric charges confined in conducting wires, only that they flow in momentum space along nodal lines. However, these frame charges are not integers, and non-Abelian signatures emerge when braiding between adjacent band nodal lines occurs, which flips the direction of the flow. In photonic systems, we discover that the photonic Γ point serves as the source or sink of such frame charge flow due to a hidden braiding induced by the often-ignored electrostatic mode at zero-frequency. We use biaxial photonic crystals as examples and show how complex nodal line configurations can be explained as the topological consequences of the frame charge flow from the Γ point to the Brillouin zone boundaries. We further designed and fabricated meta-crystals to experimentally observe these line nodes as manifestation of the non-Abelian frame charge flow.

**Main**

Topological semimetals possess gap-closing degeneracies, which are manifested as nodal points, lines, or surfaces in the three-dimensional momentum space[1-21]. These singular nodal features serve as the origin of nontrivial topology and play an important role in topological physics. The topological nature of these nodal degeneracies has been fruitfully characterized using topological invariants such as Chern numbers or quantized Berry phases defined within a single band. Very recently, the topological character of a system where multiple bands are simultaneously considered has been determined and used to characterize the nodal lines in the presence of $PT$ or $C_2T$ symmetry[22, 23]. Under such symmetry constraints, the eigenvectors can be gauged to be real. From the multiband perspective, such real eigenvectors spanning the subspace of a group of bands define an orthogonal frame, which rotates when the system Hamiltonian evolves in momentum space. Topological charges can be defined using the quantized angle of rotation of the multiband eigenvector frame around a closed loop, and such frame rotations are characterized by generalized quaternion groups[22, 24]. Distinct from single-band invariants such as Chern numbers, which are integers, quaternions are matrix-like entities that are generally noncommutative under multiplication. The multiband viewpoint thus opens the door to the exploration of non-Abelian topology in energy or frequency bands[22-38].

Despite the recent advances in investigating nontrivial multiband topology, more intriguing features of the non-Abelian topological charges remain to be discovered. Here, we propose and illustrate a series of novel topological features using ordinary photonic systems (such as photonic crystals), wherein the equifrequency surfaces intersect each other and manifest as nodal lines in momentum space. Within the context of multiband description, such nodal lines can be characterized with non-Abelian frame charges, with a well-defined sign so that an "arrow" can be assigned to nodal lines[22, 25]. Interestingly, the arrow presentation can be heuristically identified as a manner of topological charge flow along nodal lines in momentum space. We will discuss the governing rule, generating source, and observable topological consequences of such non-Abelian frame charge flow, respectively.

The rules governing the topological frame charge flow in momentum space are reminiscent of the classical electric charges flowing in a conducting wire in real space. For example, the topological charges generally remain conserved while flowing along momentum space nodal lines, and when nodal lines meet at a point, the topological charges do not accumulate so that the charge flow obeys a Kirchhoff-like rule at the junction point. And yet significant differences emerge due to the non-Abelian features of topological charges. The sign of the current on a conducting wire cannot be affected by a nearby wire, for instance, however, the direction of the topological charge flow on a nodal line can be flipped by that of an adjacent nodal line, originating from the braiding process unique to non-Abelian charges. As such, non-Abelian braiding induces the sign change of topological charge flow. When these rules are established, the complex nodal line geometries in optical systems can be rationalized and predicted.

More interestingly, we found that a hitherto unnoticed role is played by the zero-frequency electrostatic mode photonic systems. With the zero-frequency mode taken into consideration, the photonic Γ point is regarded as a triple degeneracy. Importantly, the zero-frequency band induces a hidden non-Abelian braiding, making the Γ point a source (or sink) of the non-Abelian frame charge flow. We designed a biaxial photonic crystal whose nodal lines (intersection of equifrequency surfaces) illustrate such topological charge flow. We will see that the nodal features can be explained by regarding the photonic point as a source of frame charges, which flow outwards along momentum space nodal lines to Brillouin zone (BZ) boundaries, following an analogue of Kirchhoff's law while encountering the BZ boundaries. The emergence of extra nodal lines on the BZ boundaries is enforced as the topological consequences and their experimental observation substantiates the notion of frame charge flow.

**Non-Abelian frame charge flow with and without braiding**

Nodal line is the momentum space presentation of band degeneracies that are usually protected by symmetries such as mirror or *PT*. The nodal line carries quantized Berry phase and can sometimes be viewed as a singularity that is the zero limit of a string of Berry curvature[39, 40]. In such a way, the nodal line in momentum space plays a role of confinement, where the quantized Berry phase is a result of the enclosed Berry flux by a loop encircling the nodal line. To put this narrative in a heuristic and yet quantifiable description, we can draw an analogy between topological charges and the electrical charges in conducting wire. As shown pictorially in Fig. 1(a), the magnitude of the "topological charge" in the nodal line can be defined as the quantized Berry phase accumulated along an encircling homotopy loop ($\pi_1$), which manifests as the winding of eigenvector.

However, in a multiband scenario, the quantized Berry phase becomes a non-Abelian Wilczek-Zee phase[41], and instead of considering the rotation of one single eigenvector, the rotation of the orthogonal eigenvector frame should be examined, as shown in Fig. 1(a) (right panel). Depending on whether the frame rotation is "clockwise" or "anti-clockwise", a sign can be assigned to the topological frame rotation charge, for instance, the $\pm\pi$ frame rotations are noted as the elements $q = \pm g_i$ of generalized quaternion group[24] (see Supplementary Information 1). The sign of frame rotation charge can be presented as the directed arrow on the nodal line, as shown in Fig. 1(a) (right panel).

When no braiding appears, the charge $q = \pm g_i$ remains conserved for any encircling loop moving along a nodal line due to the invariance of topology-determined Wilczek-Zee Berry phase. As such, the frame charges do not accumulate for a nodal lines crossing configuration shown in Fig. 1(b) (left panel). By incorporating the relative direction regarding the crossing point, we can define the "frame charge flow" following a rule (reminiscent of the Kirchhoff's current law) as

$$\sum_n sgn \cdot q^n = 0 \qquad (1)$$

where $n$ labels and runs over all nodal line branches, and $sgn$ characterizes the flowing direction towards or away from the junction point (in a similar manner to the definition of electric current).

Such a "no-accumulation of charge" can be illustrated by the left panel of Fig. 1(b), which shows two arrows (or frame charges $q = g_i$) flowing into a meeting point of nodal lines and then flowing out. To characterize the topology with homotopy, we encircle the nodal lines with 1D ($\pi_1$) loops. Two different ways of encircling a nodal line pair with $\pi_1$ loops are illustrated in Fig. 1(b), representing the charge multiplications of $q = \pm g_i$, which result into the $q = +1$ charge (0 frame rotation) and $q = -1$ charge ($2\pi$ frame rotation), respectively. We note that if the loop encircles two "opposite arrows" (one flowing towards and the other flowing outward from the junction point), the total charge encircled by the loop is the trivial "+1" quaternion charge, and the encircling of two "same direction arrows" (both inwards or outwards) gives non-trivial "-1" quaternion charge[22] (detail discussion in Supplementary Information 2).

However, the frame charge flow can switch sign when braiding occurs, which happens when one additional degeneracy line comes into action. The non-Abelian nature of frame rotations results in an anticommutative relation between frame charges as $g_{i+1} \cdot g_i = -g_i \cdot g_{i+1}$, which implies that a braiding between adjacent nodal lines would introduce a sign change to the frame charge or the arrow direction on a nodal line[22] (explained in Supplementary Information 2). As the presented case in Fig. 1(b) (right panel), the existence of another nodal line (the red line) at the junction can flip the arrows on blue nodal lines after they pass through the junction. We thus arrive at the arrow configuration where the arrows are rearranged as all pointing outwards (or inwards) the junction point. Such braiding point thus functions as a source or sink of the non-Abelian frame charge flow, which has not been reported before. Meanwhile, in such a configuration, any encircled nodal line pairs are with "same direction" arrows, and we arrived at the simultaneous topological charge of $q = -1$ for two $\pi_1$ loops shown in the right panel of Fig. 1(b), which is referred to as a "double -1" charge in the following for convenience.

**Generating source of frame charge flow in ordinary photonic media**

While the literature might give the impression that abstract notions such as non-Abelian topology are only relevant to very special materials, we will show here that they can be manifested in very ordinary optical materials, where the proposed topological features can be found.

In photonic systems, two electromagnetic transverse modes are generally supported due to the polarization degree of freedom. However, a zero-frequency longitudinal solution is also allowed by the Maxwell equations even though this solution is usually considered to have no consequence in wave propagation. By taking the zero-frequency solution into consideration, the photonic $\Gamma$ point becomes a threefold degeneracy point, where three eigenvectors define an orthogonal triad frame whose rotation along a $k$-space loop can be characterized by quaternions and represents the non-Abelian topological charge. In particular, we found that such intrinsic triple degeneracy at the photonic $\Gamma$ point serves as the source/sink of non-Abelian frame charge flow in momentum space that we proposed in previous section, where a hidden braiding is embedded at the $\Gamma$ point.

To reveal the non-Abelian topological features, we start from simple dielectric materials with only diagonal terms in the permittivity tensor as $\varepsilon = [\varepsilon_{xx}, \varepsilon_{yy}, \varepsilon_{zz}]$. The Maxwell equations describing such media can be encoded into a three-band Hamiltonian as (details in Supplementary Information 3)

$$H = \begin{bmatrix} \frac{k_y^2}{\varepsilon_{xx}} + \frac{k_z^2}{\varepsilon_{xx}} + \frac{\omega_{px}^2}{\varepsilon_{xx}} & -\frac{k_x k_y}{\sqrt{\varepsilon_{xx}}\sqrt{\varepsilon_{yy}}} & -\frac{k_x k_z}{\sqrt{\varepsilon_{xx}}\sqrt{\varepsilon_{zz}}} \\ -\frac{k_x k_y}{\sqrt{\varepsilon_{xx}}\sqrt{\varepsilon_{yy}}} & \frac{k_x^2}{\varepsilon_{yy}} + \frac{k_z^2}{\varepsilon_{yy}} + \frac{\omega_{py}^2}{\varepsilon_{yy}} & -\frac{k_y k_z}{\sqrt{\varepsilon_{yy}}\sqrt{\varepsilon_{zz}}} \\ -\frac{k_x k_z}{\sqrt{\varepsilon_{xx}}\sqrt{\varepsilon_{zz}}} & -\frac{k_y k_z}{\sqrt{\varepsilon_{yy}}\sqrt{\varepsilon_{zz}}} & \frac{k_x^2}{\varepsilon_{zz}} + \frac{k_y^2}{\varepsilon_{zz}} + \frac{\omega_{pz}^2}{\varepsilon_{zz}} \end{bmatrix} \quad (2)$$

whose eigenvalues give the photonic dispersion, and the three eigenvectors represent polarization states of $[\sqrt{\varepsilon_{xx}}E_x \quad \sqrt{\varepsilon_{yy}}E_y \quad \sqrt{\varepsilon_{zz}}E_z]$, which together form a rotation frame.

Taking a uniaxial material with permittivity $\varepsilon$ = [2, 2, 1] as an example, the nodal line in momentum space is shown in Fig. 1(c), where the straight nodal line in blue is formed by the intersection between the equi-frequency contours (EFCs) shown as an inset. Taking the zero-frequency solution into account (labelled as the $0^{th}$ band), the Γ point is a triple degeneracy that is marked in red. The three-band eigenpolarization frame undergoes a $2\pi$ rotation along the $\pi_1$ loop encircling the nodal line in the momentum space, as shown in Fig. 1(c). The $2\pi$ rotation corresponds to the "-1" element of the quaternion group and remains conserved when the system is adiabatically perturbed (Supplementary Information 4).

More interestingly, in a biaxial material with a permittivity tensor of $\varepsilon$ = [1, 2, 3], the nodal lines in momentum space become two lines crossing at the photonic Γ point, as shown in Fig. 1(d) and (e). As shown in these figures, there exist two loops lying on orthogonal planes (one on the horizontal plane and one on the vertical plane) that both encircle the Γ point and each gives a nontrivial topological charge of "-1", manifested as the $2\pi$ rotations of the eigenpolarization frame around the loops. The Γ point thus exhibits a "double -1" charge, which fully determines the arrow configuration as all pointing outwards (or inward). Such an arrow configuration characterizes the photonic Γ point as the generating source/sink of non-Abelian frame charge flow. Similar to the analysis in Fig. 1(c), the topological origin of the source/sink role here is due to the triple degeneracy at the photonic Γ point, which can be intuitively understood as an infinitesimal nodal line formed between the $0^{th}$ (zero-frequency mode) and $1^{st}$ bands that braids with the blue nodal lines (verified in Supplementary Information 5).

**Topological consequences manifested in the periodic Brillouin zone**

Since the photonic band of a homogeneous medium extends to infinity in k-space in the absence of a minimum length scale, the generated frame charges from the Γ point flow to infinity in momentum space. To seek observable topological signatures induced by the generating source/sink of the photonic Γ point, it is easier to consider periodic photonic crystal systems in which the "flow" of frame charges can be observed within the first Brillouin zone. In the following, we will discuss the topological consequences identified as momentum space nodal lines that are available for nearfield measurements. We note that the signature of the source/sink role of the photonic Γ point can also be observed through far-field measurements, such as by observing the polarization phase winding in conical refractions (Supplementary Information 6).

We designed an optically biaxial photonic crystal for experimental demonstration. The unit cell is shown in Fig. 2(a), where two resonators are oriented along orthogonal directions. The metal bars in the resonators have different geometrical lengths of $L_1$ and $L_2$. In the long wavelength limit, the optics of the photonic crystal can be described by the effective permittivity tensor $\varepsilon$ = [$\varepsilon_{xx}$, $\varepsilon_{yy}$, 1], where $\varepsilon_{xx} = 1 + \omega_{px}^2/(\omega^2 - \omega_{0x}^2)$ and $\varepsilon_{yy} = 1 + \omega_{py}^2/(\omega^2 - \omega_{0y}^2)$ (Supplementary Information 7). The frequencies $\omega_{px,y}$ are approximately the same due to the same periodicity of resonators arranged along the x and y directions. The resonator lengths $L_1$ and $L_2$ determine the resonance frequencies $\omega_{0x,y}$, leading to an optically biaxial material with effective $\varepsilon_{xx} \neq \varepsilon_{yy} \neq 1$ when $L_1 \neq L_2$.

The low-frequency dispersion of the photonic crystal (near Γ point) can be described using an effective Hamiltonian (Supplementary Information 7),

$$\begin{bmatrix} k_y^2 + k_z^2 + \omega_{px}^2 & -k_x k_y & -k_x k_z & -\omega_{0x}\omega_{px} & 0 \\ -k_x k_y & k_x^2 + k_z^2 + \omega_{py}^2 & -k_y k_z & 0 & -\omega_{0y}\omega_{py} \\ -k_x k_z & -k_y k_z & k_x^2 + k_y^2 & 0 & 0 \\ -\omega_{0x}\omega_{px} & 0 & 0 & \omega_{0x}^2 & 0 \\ 0 & -\omega_{0y}\omega_{py} & 0 & 0 & \omega_{0y}^2 \end{bmatrix} \begin{bmatrix} E_x \\ E_y \\ E_z \\ P_x \\ P_y \end{bmatrix} = \omega^2 \begin{bmatrix} E_x \\ E_y \\ E_z \\ P_x \\ P_y \end{bmatrix} \quad (3)$$

The eigenvalues are solved to obtain the nodal structures, and the results are shown in Fig. 2(b). Due to the existence of zero-frequency mode, the eigenvector frame rotations around the horizontal and vertical $\pi_1$ loops (indicated by the green dotted lines) can be both found to be $2\pi$ (verified in Supplementary Information 8), indicating that the Γ point is a "-1" quaternion charge from the point of view of two orthogonal planes and hence carries a "double -1" charge. Nodal lines spawned from the Γ point are each characterized by its own quaternion charge as indicated via the arrows on them in Fig. 2(b).

Going beyond the effective medium description, we now consider the frame charge flow in periodic momentum space by computing the eigenmodes of the photonic crystal (unit cell shown in Fig. 2(a)) using full-wave simulations (CST Microwave Studio). The band structures are calculated as shown in Fig. 2(c), where high symmetry positions are marked for the BZ in Fig. 2(a). The band degeneracies are retrieved as momentum space nodal lines in Fig. 2(d). The Γ point act as a source (or equivalently a sink) of non-Abelian frame charge flow. Nodal lines spawning from Γ point carry frame charges of $q = \pm g_i$ as indicated with coloured arrows on them, with the base point of the homotopy loops pined near the Γ point.

We then trace the frame charge flow in the periodic BZ of Fig. 2(d) and examine the available topological consequences. Starting from the photonic Γ point, the arrows on nodal lines labelled as #1 to #4 all go outwards directions. This is possible due to the "double -1" charge at the Γ point, where "-1" charge can be found for both the vertical loop #A and horizontal loop #B (numerically verified in Supplementary Information 9). If we follow the frame charge on nodal line #1 from Γ to the top BZ boundary ($k_z = \pi/c$ plane), we notice that the counterpart of nodal line #2 from the extended zone joins it. These lines contribute to a "-1" charge ($q = g_1 \cdot g_1 = -1$) when we consider the $\pi_1$ loop #C encircling the junction point (verified in Supplementary Information 9 and discussed in Supplementary Information 10). This fixes the arrow direction on nodal line #2', and the two arrows on nodal lines #1 and #2' are all pointing towards the junction point. To preserve the analogue Kirchhoff's law, additional nodal lines must emerge from the junction. This explains the existence of two additional nodal lines of #5 and #6 that show up on the top BZ boundary carrying frame charges that point outwards from the junction. We then see that the complex network of nodal lines has a topological reason to exist in the way it presents itself, and is recognized as the topological consequence of frame charge flow.

Furthermore, if we follow the nodal lines #5 and #6 to the right BZ boundary, we see that they were joined by nodal lines #7' and #8' from the extended zone. If we consider the $\pi_1$ loop #D in Fig. 2(d), we find that it encircles a "-1" charge, meaning that it must have two "arrows" of the same sign piercing through the area enclosed by the loop. This enforces the direction of arrows on nodal lines #5 and #7' to be the same, both flowing towards junction point. This then requires that additional nodal lines must emerge from the meeting point to satisfy the Kirchhoff-like law, which further explains the emergence of nodal lines #9 and #10' on the right BZ boundary. The Kirchhoff-like law also allows us to assign directional arrows to nodal lines #9-10 (or #11-12), as shown in Fig. 2(d). We then noticed that the arrows on #9-10 (or #11-12) are flowing in opposite directions, and yet nodal lines #9 and #10 are joined together. So, the sign of the quaternion charge on nodal line #9 must be flipped when the charge flow from the top to the bottom of the BZ on the right BZ boundary. This sign change mechanism is provided by the braiding with the red nodal ring (formed between the 2$^{nd}$ and 3$^{rd}$ bands), remembering that we have set the viewpoint (or basepoint) near the Γ point in Fig. 2(d), which makes the red nodal ring lie in front the nodal lines #9 and #10 to provide braiding. It can also be noticed that the braidings here (between red nodal ring and nodal lines #9-12) effectively serve as the sink of frame charge flow, complementary to the source at Γ point. We note again that such switching is possible because the quaternion group is non-Abelian and there is no

Abelian analogue, e.g., in electric charges. If we take a look again at the degeneracy points in Fig.2(c) and the degeneracy lines in Fig. 2(d), it is difficult to understand why they should appear in such a geometrical arrangement without using the non-Abelian topological interpretation. These nodal line configurations can thus serve as evidence of the frame charge flow, and we will provide experimental demonstration in the following.

**Experimental demonstration of non-Abelian frame charge flow**

We then fabricated the photonic crystal shown schematically in Fig. 2(a) and experimentally characterized the nodal lines on the BZ boundaries to demonstrate the frame charge flow. The samples are fabricated with printed circuit boards (PCBs) and characterized at microwave frequencies. To best present these nodal lines, we focus on the bulk band degeneracies and present the discussions related to surface modes in Supplementary Information 10 and 11.

In Fig. 3(a) and (b), we show the retrieved nodal lines on the $k_y = 0$ and $k_x = 0$ planes, respectively. These nodal lines can be experimentally characterized with the configuration shown pictorially in Fig. 3(c), where the planar resonators are fabricated on PCBs as arrays of printed metallic elements on the x – y plane. These PCBs are stacked along the z direction to construct a sample exposing the x – z or y – z surfaces.

For the nodal lines on the $k_y = 0$ plane in Fig. 3(a), the nodal degeneracies are the crossing points (marked in green) of the bulk mode EFCs, as shown in the top row of Fig. 3(d). On the x – z surface of the fabricated sample, we experimentally measured the band projections, and the projected band EFCs are shown in the bottom row of Fig. 3(d). The predicted EFCs can be clearly identified from the measured results as the outer boundaries of excited modes, verifying the nodal line on the $k_y = 0$ plane.

The nodal lines in Fig. 3(b) can be characterized by cutting the surface BZ ($k_y$ - $k_z$ plane) at discrete $k_z$ positions. In the top row of Fig. 3(e), we show the calculated band projections on the $k_y – k_z$ plane for several $k_z$-cut lines (scanned along $k_y$), where the nodal degeneracies can be identified from the projected bands marked with orange dots. In the bottom row of Fig. 3(e), we show the measured results for the projected bands for comparison to the calculation results, and we find very good agreement. The bulk band dispersions with $k_x = \pi/a$ are shown on top of the experimental results as white lines, and the crossing points are identified as orange dots.

In Fig. 4(a) and (b), we show the nodal ring on the $k_z = 0$ plane and nodal lines on the $k_z = \pm\pi/c$ plane. These degeneracies can be experimentally identified from the band projections on the $k_x$

– $k_y$ plane by cutting at different $k_x$ positions (scan along $k_y$). We show the experimental configuration in Fig. 3(c), where the PCBs are stacked along the z direction and measurements are conducted on the x – y plane.

At small values of $k_x$, the nodal ring on the $k_x$-$k_y$ plane is cut as shown in Fig. 4(a). We show the calculated band projections in the top row of Fig. 4(d). The sliced points from the nodal ring are marked as red dots in the projected bands. Drumhead surface states induced by the nodal ring can be found in the calculated results, which are marked in magenta. Their relationship to the Zak phase is discussed in Supplementary Information 10. The light cones for the background air ($\varepsilon = 1$) and substrate ($\varepsilon = 2$) are shown as cyan and red curves, respectively. We show the experimental measured results in the bottom row of Fig. 4(d). The bulk bands are shown on top of the experimental results, and the band degeneracy from the nodal ring are identified. The predicted drumhead surface states are also observed in the experimental data, as indicated with magenta dashed curves in Fig. 4(d).

For larger values of $k_x$, the nodal lines on the top (or bottom) BZ boundaries are crossed. We show the calculated results in the top row of Fig. 4(e), with the degenerate positions marked in blue. Light cones for the air and substrate are shown as green and red curves, respectively. The experimental measurements are shown in the bottom row, where the predicted degeneracies are observed as predicted and marked in blue. We note that the surface modes induced by these nodal lines are closely attached to the bulk bands, as shown in the top row of Fig. 4(e). These surface modes thus merge with the bulk band projections in the experimentally measured results.

We have now experimentally characterized all the nodal lines shown in Fig. 2(d). In particular, the observed nodal lines on the top (bottom) and side BZ boundaries serve as the topological consequence of non-Abelian frame charge flow in momentum space. These nodal lines also provide experimental evidence for the "double -1" charge at the photonic Γ point to verify the proposed generating source of frame charge flow.

In addition, a uniaxial photonic crystal can be easily achieved with the proposed photonic crystal by simply tuning $L_1 = L_2$ with the *PT* symmetry preserved. The non-Abelian topology predicted nodal lines thus remain intact at the top and side surfaces of the BZ being robust to geometrical modulations. Discussions and experimental characterizations of the uniaxial photonic crystal are provided in Supplementary Information 12 and 13. Furthermore, by reducing the *PT* symmetry to $C_2T$ symmetries, the non-Abelian frame charges are confined to the $C_2T$-invariant planes, but the emerging Berry flux exhibits interestingly similar flowing behaviour as the original frame charges. The detailed discussion is provided in Supplementary Information 14.

In conclusion, we proposed the flowing behaviour of non-Abelian frame charges along nodal lines in momentum space. Some of the behaviours can be viewed as an analogue of electric charges flowing along conducting wires in real space, but the possibility of sign flipping (as induced by the passing over of an adjacent nodal line) is uniquely non-Abelian. Non-Abelian band topology manifests itself in simple photonic systems, and we showed that the photonic Γ point in ordinary optical media are the source or sink of non-Abelian frame charge flow. We fabricated a biaxial photonic crystal and experimentally characterized the degeneracy features to demonstrate the frame charge flow in momentum space. Our results have demonstrated the flowing behaviour of non-Abelian frame charges and shed new light on the fundamental understanding of photonic bands, which could inspire further exploration of the novel non-Abelian physics in ordinary materials.

**Acknowledgments**

We would like to thank Dr. Ruo-Yang Zhang, Dr. Biao Yang, and Prof. Shuang Zhang for very helpful discussion. This work is supported by Research Grants Council of Hong Kong through grants 16307821, AoE/P-502/20, 16310420, 16307621 and by KAUST20SC01.

## Figures

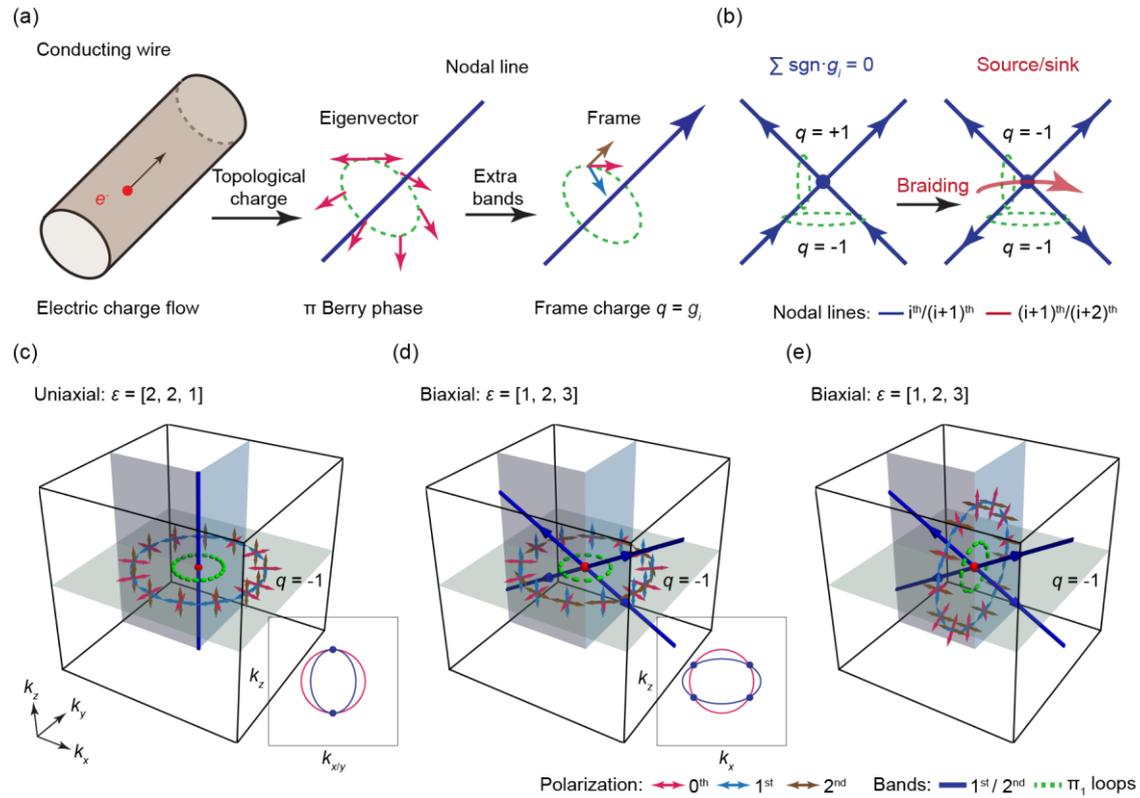

Fig. 1. Non-Abelian frame charge flow and the source at the triple-degenerate Γ point. (a) The electric charge moving in a conducting wire is a real space counterpart of the topological charge (quantized Berry phase or frame rotation) moving along the nodal line in momentum space. (b) When two nodal lines cross each other, the frame charges do not accumulate at the joint point and follow an analogue Kirchhoff's current law. When "braiding" with another nodal line occurs, a sourceless junction point transforms into a source or sink of frame charges with the "all-out" or "all-in" arrow configuration. (c) The photonic Γ point is a triple degeneracy by taking the zero-frequency mode into consideration, as marked in red. The nodal line (the intersection of the EFCs from inset) of a uniaxial material gives rise to $2\pi$ rotation of the eigenpolarization frame, corresponding to the "-1" quaternion charge. (d-e) Nodal lines of the biaxial material form as the cross structure in momentum space. A hidden braiding buried in the triple degeneracy gives rise to the source/sink of non-Abelian frame charge flow at the photonic Γ point. The two $\pi_1$ loops in the horizontal and vertical planes both give $2\pi$ frame rotations, which reveals the "double -1" charge that generates the frame charge flow.

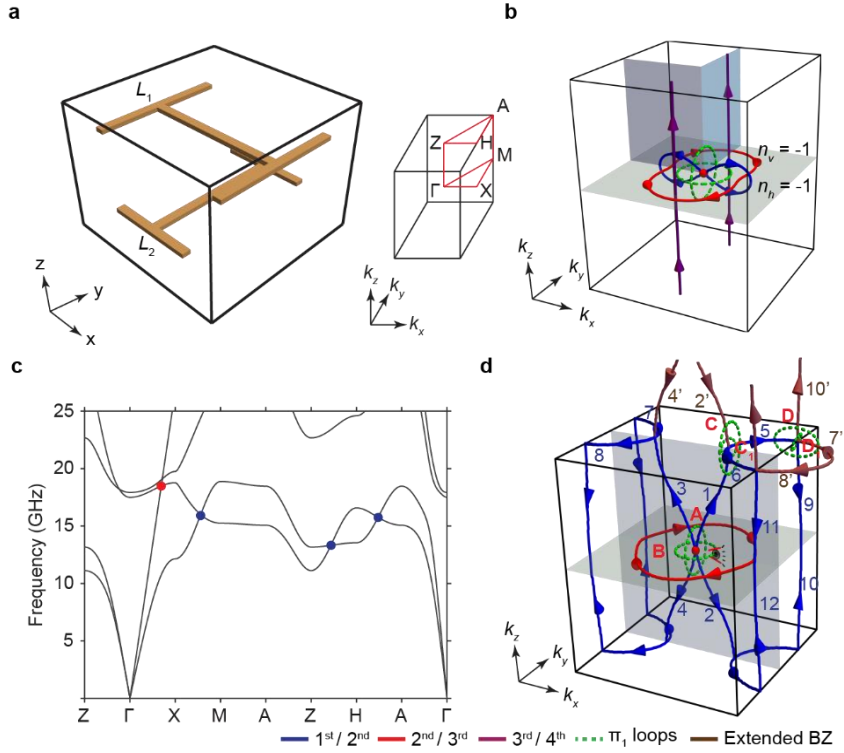

Fig. 2. Non-Abelian frame charge flow in biaxial photonic crystal. (a) Unit cell containing two H-shaped resonators with $L_1 = 3$ mm and $L_2 = 2$ mm. The lattice constants along the x, y, and z directions are $a = 4$ mm, $b = 4$ mm, and $c = 3.2$ mm, respectively. The BZ is shown in the right inset. (b) Nodal structures from the effective Hamiltonian calculation. The Γ point carries a "double -1" charge, which can be viewed as the source of a non-Abelian frame charge flow, as indicated with directed arrows in colour. Adopted parameters are $\omega_{px,y} = 1$, $\omega_{0x} = 0.4$, and $\omega_{0y} = 1.6$. (c) Full wave band structures calculated for the photonic crystal. Degeneracy points are marked in colour. (d) Retrieved nodal structure from photonic crystal band dispersions. Brown nodal lines are degeneracies between the 1st and 2nd bands but join from the extended BZ. Green circles indicate $\pi_1$ loops characterizing a "-1" (A, B, C and D) or "+1" element of the generalized quaternion group ($C_1$ and $D_1$). Nodal lines #5-12 can be viewed as topological consequences of the non-Abelian frame charge flow originating from the source at the photonic Γ point.

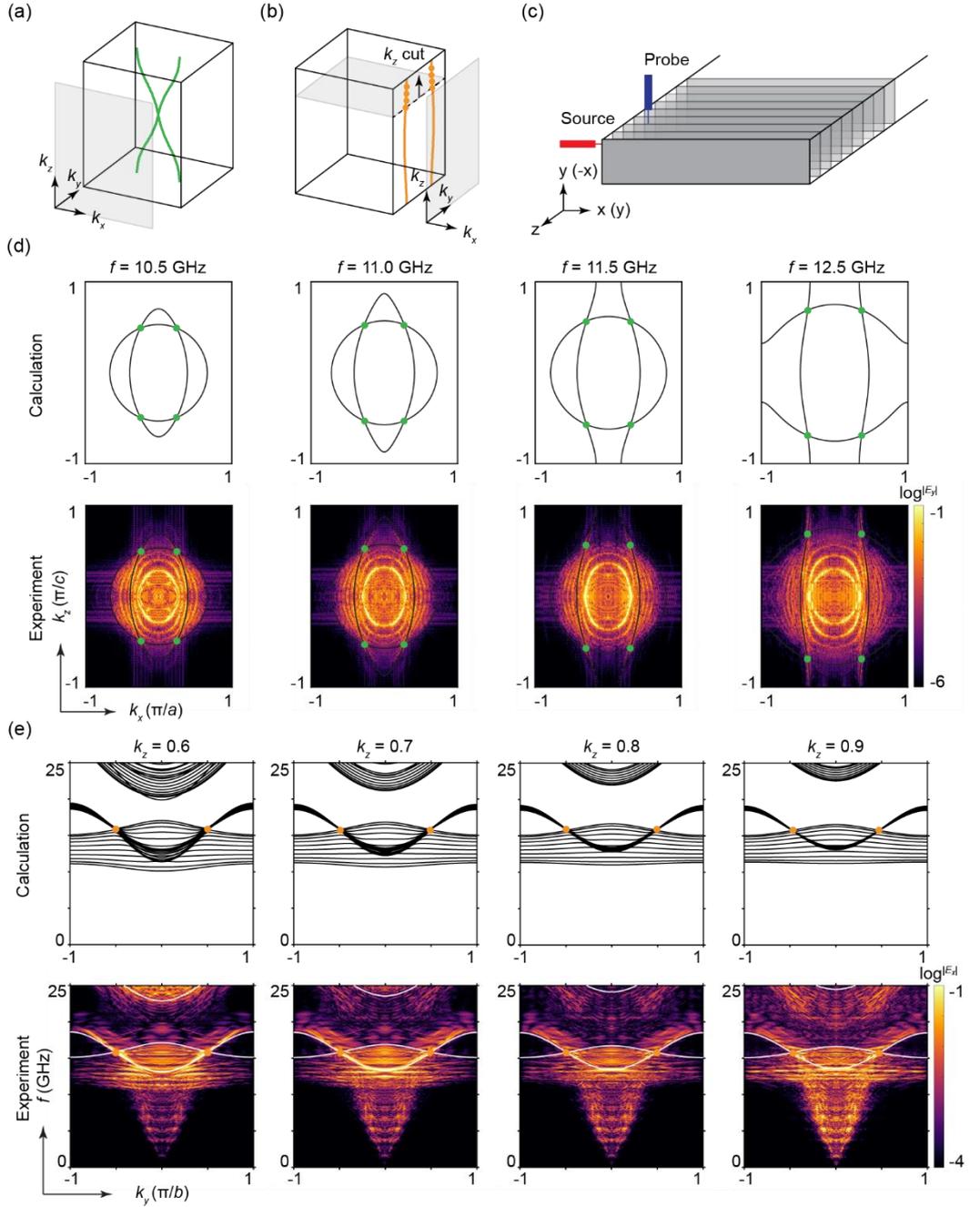

Fig. 3. Characterization of nodal lines on the x – z and y – z surfaces of the photonic crystal. (a) The nodal line in green is located on the $k_y = 0$ plane and can be measured from the x – z surface. (b) The nodal lines at the $k_x = \pm\pi/a$ boundaries can be characterized from the y – z surface. (c) The experimental configuration for side surface measurements. The photonic crystal PCBs are stacked horizontally (along the x or y direction, 80×80×10 units), and the measurements are conducted on the x – z or y – z surface. The source and probe antennas are arranged as indicated. (d) The green nodal line in (a) represents the EFC crossings shown for different frequencies in the top row. Experimentally measured results are shown in the bottom

row. (e) Calculated band projection on the $k_y - k_z$ plane with $k_z$ fixed at different values (scan along $k_y$). Band degeneracies are marked in orange. Experimentally measured results are shown in the bottom row, and bulk bands are plotted as white curves.

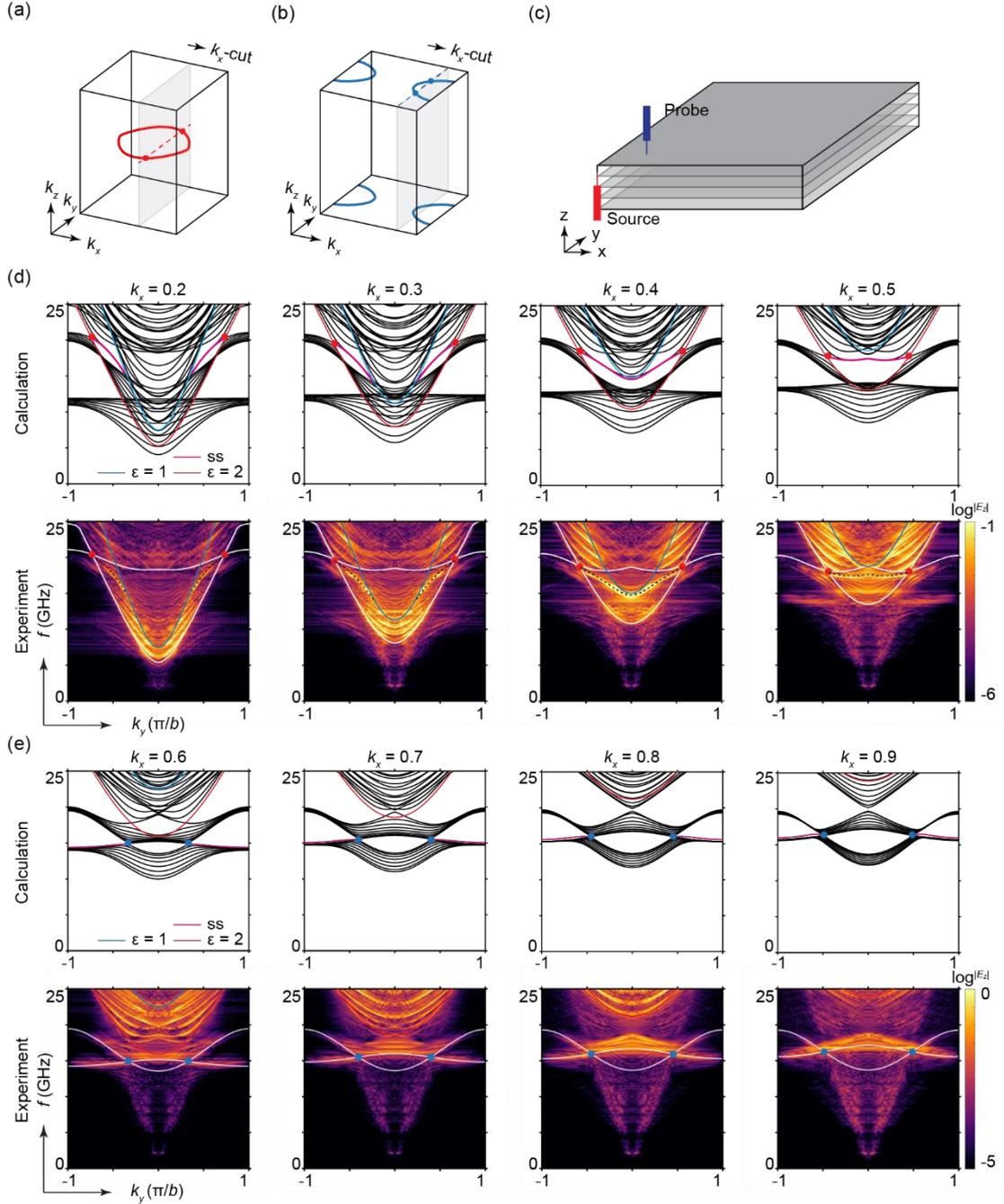

Fig. 4. Characterization of nodal lines on the x – y surface. (a) The plane at small $k_x$ value ($k_x$ = *constant*) cuts the nodal ring at two nodal points. (b) For large values of $k_x$, the blue nodal lines are intersected by the $k_x$=*constant* plane. (c) Measurement configuration, where PCBs comprising the photonic crystal are stacked along the z direction (80×80×10 units). (d) Simulation results for the projected bands at small $k_x$ values (scanned along $k_y$) are shown in the first row. Experimentally measured band projections are shown in the second row, with computed band dispersions overlaid as white curves, and the blue line is the light cone. Drumhead surface modes are also experimentally observed. (e) Calculated band projections at large $k_x$ values are shown in the top row, and the degeneracy nodes are marked in blue. The experimentally measured results are shown in the bottom row, and the computed bulk bands are plotted as white curves.